\documentclass[12pt]{article}
\usepackage{myart}

\renewcommand{\theequation}{\arabic{section}.\arabic{equation}}
\normalbaselines
\input tcilatex.tex

\begin{document}

\author{Yuri A. Rylov}
\title{What object does the wave function describe?}
\date{Institute for Problems in Mechanics, Russian Academy of Sciences \\
101-1,Vernadskii Ave., Moscow, 119526, Russia \\
email: rylov@ipmnet.ru\\
Web site: {$http://rsfq1.physics.sunysb.edu/\symbol{126}rylov/yrylov.htm$}\\
or mirror Web site: {$http://gasdyn-ipm.ipmnet.ru/\symbol{126}%
rylov/yrylov.htm$}}
\maketitle

\begin{abstract}
It is shown that the wave function describes the state of the statistical
ensemble $\mathcal{E}\left[ \mathcal{S}\right] $ of individual particles, or
the statistical average particle $\left\langle \mathcal{S}\right\rangle $.
This result follows from the fact that in the classical limit $\hbar =0$ the
Schr\"{o}dinger equation turns to the dynamic equations for the statistical
ensemble of classical particles. The statement that the wave function
describes the state of an individual particle is incompatible with the
quantum mechanics formalism. It is shown that the statistical interpretation
of quantum mechanics is a corollary of the fact, that the QM formalism is
the technique of the statistical ensemble description, restricted by
constraints of the dynamic equation linearity.
\end{abstract}

\section{Introduction}

By definition the quantum system $\mathcal{S}_{\mathrm{q}}$ is a continuous
dynamic system, whose state is described by the wave function. There are two
different opinions about the quantum system $\mathcal{S}_{\mathrm{q}}$. Is
the quantum system $\mathcal{S}_{\mathrm{q}}$ an individual dynamic (or
stochastic) system $\mathcal{S}$, or is it a statistical ensemble $\mathcal{E%
}\left[ \mathcal{S}\right] $ of dynamic (or stochastic) systems $\mathcal{S}$%
? In the conventional Copenhagen interpretation \cite{H55} of quantum
mechanics $\mathcal{S}_{\mathrm{q}}=\mathcal{S}$, whereas in another
version, what is known as statistical interpretation \cite
{B76,B70,B98,FP00,BYZ94}, $\mathcal{S}_{\mathrm{q}}=\mathcal{E}\left[ 
\mathcal{S}\right] $. Discussion on this subject lasts many years on the
verbal level without a visible progress. There is a lot of papers on this
subject \cite{P64}-\cite{G90}, where the discussion is produced on the
qualitative (verbal) level without any connection with the formalism of
quantum mechanics. All authors believe in principles and formalism of the
conventional quantum mechanics, but their interpretation of quantum
mechanics is different, in general. There is a wide spectrum of opinions,
because different authors use different primary suppositions, and it is
impossible to decide which of them are valid. It is useless to consider and
compare opinions of different authors, made on the verbal level, if one can
solve the question on the mathematical level, using strictly defined
concepts and quantities.

The question, what the wave function does describe, is not a question of a
belief. This question can be solved on the foundation of the correspondence
principle and the quantum mechanics (QM) formalism. If in the classical
limit the quantum system $\mathcal{S}_{\mathrm{q}}$ turns into a classical
dynamic system $\mathcal{S}$, the quantum system $\mathcal{S}_{\mathrm{q}}$
is an individual dynamic system $\mathcal{S}$. If in the classical limit the
quantum system $\mathcal{S}_{\mathrm{q}}$ turns into a statistical ensemble $%
\mathcal{E}\left[ \mathcal{S}\right] $ of classical dynamic systems, the
quantum system $\mathcal{S}_{\mathrm{q}}$ is a statistical ensemble of
(stochastic) systems. In the present paper we try to solve this question
mathematically on the basis of the quantum mechanics formalism without
application of any additional suppositions.

First of all, we shall distinguish between the individual quantum-stochastic
particle (system) $\mathcal{S}_{\mathrm{st}}$ and the quantum particle
(system) $\mathcal{S}_{\mathrm{q}}$, whose state is described by the wave
function $\psi $. The particle $\mathcal{S}_{\mathrm{st}}$ is a discrete
stochastic system in the sense that it has a finite number (six) of the
freedom degrees, and there exist no dynamic equations for it. On the
contrary, the quantum particle (system) $\mathcal{S}_{\mathrm{q}}$ is a
continuous dynamic system in the sense that it has infinite number of the
freedom degrees, and there are dynamic equations for the wave function which
describes the state of $\mathcal{S}_{\mathrm{q}}$. We shall show that in the
limit $\hbar \rightarrow 0$ the quantum particle (system) $\mathcal{S}_{%
\mathrm{q}}$ turns into the statistical ensemble $\mathcal{E}\left[ \mathcal{%
S}_{\mathrm{cl}}\right] $ of classical particles $\mathcal{S}_{\mathrm{cl}}$%
. Strictly, we cannot say anything about transformation of the
quantum-stochastic particle $\mathcal{S}_{\mathrm{st}}$ at $\hbar
\rightarrow 0$, because we have no information about $\mathcal{S}_{\mathrm{st%
}}$ (there are no dynamic equations for $\mathcal{S}_{\mathrm{st}}$). We may
suppose that at $\hbar \rightarrow 0$ the quantum-stochastic particle $%
\mathcal{S}_{\mathrm{st}}$ turns into the deterministic classical particle $%
\mathcal{S}_{\mathrm{cl}}$, but we cannot use this supposition, because the
quantum mechanics (QM) formalism does not deal with $\mathcal{S}_{\mathrm{st}%
}$. It deals only with $\mathcal{S}_{\mathrm{q}}$. We can only show that
this supposition is compatible with the formalism of quantum mechanics.

Thus, we are going to prove that 
\begin{equation}
\text{at }\hbar \rightarrow 0,\qquad \mathcal{S}_{\mathrm{q}}\rightarrow 
\mathcal{E}\left[ \mathcal{S}_{\mathrm{cl}}\right]  \label{a1.1a}
\end{equation}
The alternative statement has the form 
\begin{equation}
\text{at }\hbar \rightarrow 0,\qquad \mathcal{S}_{\mathrm{q}}\rightarrow 
\mathcal{S}_{\mathrm{cl}}  \label{a1.1b}
\end{equation}
We have a pure mathematical problem. The systems $\mathcal{S}_{\mathrm{q}}$
and $\mathcal{E}\left[ \mathcal{S}_{\mathrm{cl}}\right] $ are well-known
continuous dynamic systems. The system $\mathcal{S}_{\mathrm{cl}}$ is the
known discrete dynamic system. Dependence of $\mathcal{S}_{\mathrm{q}}$ on
the parameter $\hbar $ is analytical. We can verify which one of relations (%
\ref{a1.1a}), (\ref{a1.1b}) is valid. This will be a mathematical solution
of the considered problem.

One can see at once, that transformation (\ref{a1.1b}) of the continuous
dynamic system $\mathcal{S}_{\mathrm{q}}$ into the discrete dynamic system $%
\mathcal{S}_{\mathrm{cl}}$ seems to be rather problematic, although we
cannot ignore completely the possibility of such a degeneration of the
continuous dynamic system. Transformation (\ref{a1.1a}) of the continuous
dynamic system $\mathcal{S}_{\mathrm{q}}\ $into continuous $\mathcal{E}\left[
\mathcal{S}_{\mathrm{cl}}\right] $ seems to be more verisimilar.

For the free nonrelativistic particle we have the following expressions for
the actions 
\begin{equation}
\mathcal{S}_{\mathrm{q}}:\qquad \mathcal{A}_{\mathcal{S}_{\mathrm{q}}}\left[
\psi ,\psi ^{\ast }\right] =\int \left\{ \frac{i\hbar }{2}\left( \psi ^{\ast
}\partial _{0}\psi -\partial _{0}\psi ^{\ast }\cdot \psi \right) -\frac{%
\hbar ^{2}}{2m}\mathbf{\nabla }\psi ^{\ast }\mathbf{\nabla }\psi \right\} dtd%
\mathbf{x}  \label{a1.2}
\end{equation}
where $\psi =\psi \left( t,\mathbf{x}\right) $ is a complex one-component
wave function, $\psi ^{\ast }=\psi ^{\ast }\left( t,\mathbf{x}\right) $ is
the complex conjugate to $\psi $, and $m$ is the particle mass. 
\begin{equation}
\mathcal{S}_{\mathrm{cl}}:\qquad \mathcal{A}_{\mathcal{S}_{\mathrm{cl}}}%
\left[ \mathbf{x}\right] =\int \frac{m}{2}\left( \frac{d\mathbf{x}}{dt}%
\right) ^{2}dt  \label{a1.21}
\end{equation}
where $\mathbf{x}=\left\{ x^{1}\left( t\right) ,x^{2}\left( t\right)
,x^{3}\left( t\right) \right\} $ and $m$ is the particle mass.

By definition the pure statistical ensemble $\mathcal{E}\left[ \mathcal{S}_{%
\mathrm{cl}}\right] $ of classical particles $\mathcal{S}_{\mathrm{cl}}$ is
such a statistical ensemble, where the distribution function $F\left( 
\mathbf{x},\mathbf{p}\right) $ has the form 
\begin{equation}
F\left( t,\mathbf{x},\mathbf{p}\right) =\rho \left( t,\mathbf{x}\right)
\delta \left( \mathbf{p}-\mathbf{P}\left( t,\mathbf{x}\right) \right)
\label{a1.21a}
\end{equation}
where $\rho $ and $\mathbf{P}$ are function of only $t,\mathbf{x}$.

The action for the pure statistical ensemble $\mathcal{E}\left[ \mathcal{S}_{%
\mathrm{cl}}\right] $ of classical particles $\mathcal{S}_{\mathrm{cl}}$ can
be represented in the form \cite{R2002} 
\begin{equation}
\mathcal{E}\left[ \mathcal{S}_{\mathrm{cl}}\right] :\qquad \mathcal{A}_{%
\mathcal{E}\left[ \mathcal{S}_{\mathrm{cl}}\right] }\left[ \mathbf{x}\right]
=\int \frac{m}{2}\left( \frac{d\mathbf{x}}{dt}\right) ^{2}dtd\mathbf{\xi }
\label{a1.22}
\end{equation}
where $\mathbf{x}=\left\{ x^{1}\left( t,\mathbf{\xi }\right) ,x^{2}\left( t,%
\mathbf{\xi }\right) ,x^{3}\left( t,\mathbf{\xi }\right) \right\} $.
Parameters $\mathbf{\xi =}\left\{ \xi _{1},\xi _{2},\xi _{3}\right\} $ label
elements (particles) of the statistical ensemble $\mathcal{E}\left[ \mathcal{%
S}_{\mathrm{cl}}\right] $. The action (\ref{a1.22}) is a sum (integral) of
actions (\ref{a1.21}). We see from (\ref{a1.21}) and (\ref{a1.22}) that
there is one-to-one correspondence between the Lagrange function of $%
\mathcal{S}_{\mathrm{cl}}$ and that of $\mathcal{E}\left[ \mathcal{S}_{%
\mathrm{cl}}\right] $.

The mathematical statement of the problem is very simple and evident. But as
far as we know, it was not considered and solved in such a form. Why? There
are two mathematical problems which are to be solved, before a comparison of
the actions (\ref{a1.2}) and (\ref{a1.22}) appears to be possible. To
compare actions (\ref{a1.2}) and (\ref{a1.22}) we are to transform them to
the same independent and dependent variables.

The first problem is connected with the description of the action (\ref
{a1.22}) in terms of the wave function. The action (\ref{a1.22}) is written
in the Lagrangian (independent) coordinates $t,\mathbf{\xi }$\textbf{. }It
can be easily written in the Eulerian (independent) coordinates $t,\mathbf{x}
$\textbf{, }and the dynamic system $\mathcal{E}\left[ \mathcal{S}_{\mathrm{cl%
}}\right] $ turns to some fluid without pressure. It is well known \cite{M26}
that the Schr\"{o}dinger equation 
\begin{equation}
i\hbar \partial _{0}\psi =-\frac{\hbar ^{2}}{2m}\mathbf{\nabla }^{2}\psi
,\qquad -i\hbar \partial _{0}\psi ^{\ast }=-\frac{\hbar ^{2}}{2m}\mathbf{%
\nabla }^{2}\psi ^{\ast }  \label{a1.2a}
\end{equation}
generated by the action (\ref{a1.2}) also can be written in the hydrodynamic
form. It is sufficient to make the change of variables 
\begin{equation}
\psi =\sqrt{\rho }e^{i\varphi },\qquad \psi ^{\ast }=\sqrt{\rho }%
e^{-i\varphi },  \label{b1.16a}
\end{equation}
to substitute (\ref{b1.16a}) in (\ref{a1.2a}) and to separate real and
imaginary part of the equation. We obtain two expressions for $\partial
_{0}\rho $ and $\partial _{0}\varphi $. Taking gradient $\mathbf{\nabla }%
\partial _{0}\varphi $ and introducing the velocity $\mathbf{v=}$ $\mathbf{%
\nabla }\varphi $, we obtain four hydrodynamic equations for four dependent
variables $\rho $, $\mathbf{v}$. Thus, to pass from the description of $%
\mathcal{S}_{\mathrm{q}}$ in terms of the wave function to the
hydrodynamical description, we need to differentiate dynamic equations. It
means that for transition from the hydrodynamic description to the
description in terms of the wave function, we are to integrate hydrodynamic
equations. Problem of integration of hydrodynamic equations is rather simple
in the case of irrotational flow, but it is a difficult problem in the
general case of the rotational flow. It means that for comparison of the
actions (\ref{a1.2}) and (\ref{a1.22}), we are to integrate dynamic
equations for the ideal fluid without a pressure in the general case. The
general integration of hydrodynamic equations, which is accompanied by
appearance of three arbitrary function of three arguments, is the first
mathematical problem. This problem has been solved only in the end of
eighties \cite{R89}. Until this solution the comparison of the actions (\ref
{a1.2}) and (\ref{a1.22}) was impossible, because the dynamic equations,
generated by the action (\ref{a1.2}) are dynamic equations for integrals of
hydrodynamic equations.

These integrals have the form \cite{R99} 
\begin{equation}
\mathbf{p}=m\mathbf{v}=\frac{b}{2}\left( \mathbf{\nabla }\varphi +g^{\alpha
}\left( \mathbf{\xi }\right) \mathbf{\nabla }\xi _{\alpha }\right) ,
\label{b1.7}
\end{equation}
where $\varphi $ is a new dependent variable, $\mathbf{\xi =}\left\{ \xi
_{1},\xi _{2},\xi _{3}\right\} $ are Lagrangian coordinates considered as
functions of $t,\mathbf{x}$; $b\neq 0$ is an arbitrary constant (the
integration constant) and $g^{\alpha }\left( \mathbf{\xi }\right) $, \ $%
\alpha =1,2,3$ are arbitrary functions (result of integration). If $b=0$,
integrals (\ref{b1.7}) degenerate into zero, and the description in terms of
Lagrangian coordinates $\mathbf{\xi }$ disappears.

The function $\psi $ is constructed of the variable $\varphi $, the fluid
density $\rho $ and the Lagrangian coordinates $\mathbf{\xi }$, considered
as functions of $\left( t,\mathbf{x}\right) $, as follows \cite{R99}. The $n$%
-component complex function $\psi =\{\psi _{\alpha }\},\;\;\alpha
=1,2,\ldots ,n$ is defined by the relations 
\begin{equation}
\psi _{\alpha }=\sqrt{\rho }e^{i\varphi }u_{\alpha }(\mathbf{\xi }),\qquad
\psi _{\alpha }^{\ast }=\sqrt{\rho }e^{-i\varphi }u_{\alpha }^{\ast }(%
\mathbf{\xi }),\qquad \alpha =1,2,\ldots ,n,  \label{s1.1}
\end{equation}
\begin{equation}
\psi ^{\ast }\psi \equiv \sum_{\alpha =1}^{n}\psi _{\alpha }^{\ast }\psi
_{\alpha },  \label{s1.2}
\end{equation}
where (*) means the complex conjugate. The quantities $u_{\alpha }(\mathbf{%
\xi })$, $\alpha =1,2,\ldots ,n$ are functions of only variables $\mathbf{%
\xi }$, and satisfy the relations 
\begin{equation}
-{\frac{i}{2}}\sum_{\alpha =1}^{n}\left( u_{\alpha }^{\ast }\frac{\partial
u_{\alpha }}{\partial \xi _{\beta }}-\frac{\partial u_{\alpha }^{\ast }}{%
\partial \xi _{\beta }}u_{\alpha }\right) =g^{\beta }(\mathbf{\xi }),\qquad
\beta =1,2,3,\qquad \sum_{\alpha =1}^{n}u_{\alpha }^{\ast }u_{\alpha }=1.
\label{s5.5}
\end{equation}
The number $n$ is such a natural number that equations (\ref{s5.5}) admit a
solution. In general, $n$ depends on the form of the arbitrary integration
functions $\mathbf{g}=\{g^{\beta }(\mathbf{\xi })\}$, $\beta =1,2,3$. The
functions $\mathbf{g}$ determine vorticity of the fluid flow. If $\mathbf{g}%
=0$, equations (\ref{s5.5}) have the solution $u_{1}=1$, $u_{\alpha }=0$, \ $%
\alpha =2,3,...n$. In this case the function $\psi $ may have one component,
and the fluid flow is irrotational. The function $\psi $ has the form (\ref
{b1.16a}) and it does not depend on the Lagrangian coordinates $\mathbf{\xi }
$. To compare the dynamic system (\ref{a1.22}) and the dynamic system (\ref
{a1.2}) with $\hbar \rightarrow 0$, we suppose that the function $\psi $,
constructed by relations (\ref{b1.7}) - (\ref{s5.5}) for the dynamic system (%
\ref{a1.22}), is the wave function, which appears in the action (\ref{a1.2})
as a dependent variable.

The second problem is connected with the limit $\hbar \rightarrow 0$ in the
action (\ref{a1.2}). We cannot transit to this limit, because the action (%
\ref{a1.2}) vanishes and the quantum dynamic system $\mathcal{S}_{\mathrm{q}%
} $ degenerates. The dynamic system (\ref{a1.2}) is a quantum system. The
quantum systems form a special class of dynamic systems, which satisfy the
QM principles, i.e. dynamic equations (\ref{a1.2a}) are linear. The state of
any quantum system is described by a special complex dependent variable $%
\psi $, known as the wave function. The wave function is considered to be a
specific quantum object, which may be considered as a vector in the Hilbert
space. Transformations of the wave functions (state vectors) may be only
linear. At the nonlinear transformation of the dependent variable (wave
function) the dynamic system $\mathcal{S}_{\mathrm{q}}$ remains to be the
same dynamic system, but it ceases to be the quantum system, because dynamic
equations become nonlinear and dynamic system ceases to satisfy the QM
principles. In other words, the property of a dynamic system to be a quantum
system is connected with some constraints on dependent variables.

Dynamic system (\ref{a1.2a}) is the quantum system for any value of the
parameter $\hbar $ except for $\hbar =0$. Let $\mathcal{SQ}$ be the set of
quantum systems $Q_{\hbar }=\mathcal{S}_{\mathrm{q}}$ for all $\hbar \neq 0$%
. We may say that $Q_{\hbar }\in \mathcal{SQ}$, for all values of the
parameter $\hbar $ except for $\hbar =0$. Let $\mathcal{SD}$ be the set of
all dynamic systems. Then $\mathcal{SQ\subset SD}$. We expect that for $%
\hbar \rightarrow 0$ the quantum system $Q_{\hbar }\in \mathcal{SQ}$ turns
to the classical (non-quantum) dynamic system $\mathcal{C}$

\begin{equation}
\mathcal{C}=\lim_{\hbar \rightarrow 0}Q_{\hbar },\qquad \mathcal{C}\notin 
\mathcal{SQ},\qquad \mathcal{C}\in \mathcal{SD}  \label{a2.1}
\end{equation}
To realize the limit (\ref{a2.1}) in the action (\ref{a1.2a}) we are to go
outside the set $\mathcal{SQ}$ of quantum systems into the set $\mathcal{SD}$
of all dynamic systems, because the limiting classical dynamic system $%
\mathcal{C}$ does not belong to $\mathcal{SQ}$.

\section{Transformations of the action for the quantum particle $\mathcal{S}%
_{\mathrm{q}}.$}

With the action $\mathcal{A}_{\mathcal{S}_{\mathrm{q}}}\left[ \psi ,\psi
^{\ast }\right] $, determined by the relation (\ref{a1.2}) the following
canonical quantities are associated: 
\begin{equation}
j^{k}=\left\{ \rho ,\mathbf{j}\right\} =\frac{i}{\hbar }\left( \frac{%
\partial \mathcal{L}}{\partial \left( \partial _{k}\psi ^{\ast }\right) }%
\psi ^{\ast }-\frac{\partial \mathcal{L}}{\partial \left( \partial _{k}\psi
\right) }\psi \right) ,\qquad \partial _{k}\equiv \frac{\partial }{\partial
x^{k}},\qquad k=0,1,2,3  \label{a1.3}
\end{equation}
\begin{equation}
T_{l}^{k}=\frac{\partial \mathcal{L}}{\partial \left( \partial _{k}\psi
^{\ast }\right) }\partial _{l}\psi ^{\ast }+\frac{\partial \mathcal{L}}{%
\partial \left( \partial _{k}\psi \right) }\partial _{l}\psi -\delta _{l}^{k}%
\mathcal{L},\qquad k,l=0,1,2,3  \label{a1.4}
\end{equation}
where $\mathcal{L}$ is the Lagrangian density for the action (\ref{a1.2}) 
\begin{equation}
\mathcal{L}=\frac{i\hbar }{2}\left( \psi ^{\ast }\partial _{0}\psi -\partial
_{0}\psi ^{\ast }\cdot \psi \right) -\frac{\hbar ^{2}}{2m}\mathbf{\nabla }%
\psi ^{\ast }\mathbf{\nabla }\psi  \label{a1.4a}
\end{equation}
We have 
\begin{equation}
\rho =\psi ^{\ast }\psi ,\qquad \mathbf{j}=-\frac{i\hbar }{2m}\left( \psi
^{\ast }\mathbf{\nabla }\psi -\mathbf{\nabla }\psi ^{\ast }\cdot \psi \right)
\label{a1.5}
\end{equation}
\begin{equation}
T_{0}^{0}=\frac{\hbar ^{2}}{2m}\mathbf{\nabla }\psi ^{\ast }\mathbf{\nabla }%
\psi ,\qquad T_{\alpha }^{0}=\frac{i\hbar }{2}\left( \psi ^{\ast }\partial
_{\alpha }\psi -\partial _{\alpha }\psi ^{\ast }\cdot \psi \right)
=mj_{\alpha }=-mj^{\alpha },\qquad \alpha =1,2,3  \label{a1.6}
\end{equation}
\begin{equation}
T_{0}^{\alpha }=-\frac{\hbar ^{2}}{2m}\left( \partial _{\alpha }\psi
\partial _{0}\psi ^{\ast }+\partial _{\alpha }\psi ^{\ast }\partial _{0}\psi
\right) =-\frac{i\hbar ^{3}}{4m^{2}}\left( \partial _{\alpha }\psi ^{\ast }%
\mathbf{\nabla }^{2}\psi -\mathbf{\nabla }^{2}\psi ^{\ast }\partial _{\alpha
}\psi \right) ,\qquad \alpha =1,2,3  \label{a1.7}
\end{equation}
\begin{equation}
T_{\beta }^{\alpha }=\frac{\hbar ^{2}}{2m}\partial _{\alpha }\psi \partial
_{\beta }\psi ^{\ast }+\frac{\hbar ^{2}}{2m}\partial _{\alpha }\psi ^{\ast
}\partial _{\beta }\psi -\delta _{\beta }^{\alpha }\mathcal{L},\qquad \alpha
,\beta =0,1,2,3  \label{a1.8}
\end{equation}

Note that this association is produced on the dynamical level, i.e.
independently of the QM principles. Meaning of the canonical quantities $%
\rho $, $\mathbf{j}$, $T_{l}^{k}$ can be obtained by means of the
correspondence principle from the meaning of these quantities in the
classical limit $\hbar \rightarrow 0$.

It is easy to see, that we may not set $\hbar =0$ in the action (\ref{a1.2})
and in the expression for the canonical quantities (\ref{a1.5}) - (\ref{a1.8}%
), because in this case we obtain no description of the dynamical system $%
\mathcal{S}_{\mathrm{q}}$. Before transition to the classical description we
should transform the phase of the wave function $\psi $. We make the change
of variables 
\begin{equation}
\psi \rightarrow \Psi _{b}=\left| \psi \right| \exp \left( \frac{\hbar }{b}%
\log \frac{\psi }{\left| \psi \right| }\right) ,\qquad \psi =\left| \Psi
_{b}\right| \exp \left( \frac{b}{\hbar }\log \frac{\Psi _{b}}{\left| \Psi
_{b}\right| }\right)   \label{a1.9}
\end{equation}
where $b\neq 0$ is some real constant. After this change of variables the
action (\ref{a1.2}) turns into 
\[
\mathcal{S}_{\mathrm{q}}:\qquad \mathcal{A}_{\mathcal{S}_{\mathrm{q}}}\left[
\Psi _{b},\Psi _{b}^{\ast }\right] =\int \left\{ \frac{ib}{2}\left( \Psi
_{b}^{\ast }\partial _{0}\Psi _{b}-\partial _{0}\Psi _{b}^{\ast }\cdot \Psi
_{b}\right) -\frac{b^{2}}{2m}\mathbf{\nabla }\Psi _{b}^{\ast }\mathbf{\nabla 
}\Psi _{b}\right. 
\]
\begin{equation}
-\left. \frac{\hbar ^{2}-b^{2}}{8m}\left( \mathbf{\nabla }\left| \Psi
_{b}\right| \right) ^{2}\right\} dtd\mathbf{x}  \label{a1.10}
\end{equation}

The dynamic equation takes the form 
\begin{equation}
ib\partial _{0}\Psi _{b}=-\frac{b^{2}}{2m}\mathbf{\nabla }^{2}\Psi _{b}-%
\frac{\hbar ^{2}-b^{2}}{8m}\left( \frac{\left( \mathbf{\nabla }\rho \right)
^{2}}{\rho ^{2}}+2\mathbf{\nabla }\frac{\mathbf{\nabla }\rho }{\rho }\right)
\Psi _{b},\qquad \rho \equiv \Psi _{b}^{\ast }\Psi _{b}  \label{a1.11}
\end{equation}

The relations (\ref{a1.5}), (\ref{a1.6}) take the form 
\begin{equation}
\rho =\Psi _{b}^{\ast }\Psi _{b},\qquad \mathbf{j}=-\frac{ib}{2m}\left( \Psi
_{b}^{\ast }\mathbf{\nabla }\Psi _{b}-\mathbf{\nabla }\Psi _{b}^{\ast }\cdot
\Psi _{b}\right)  \label{a1.12}
\end{equation}
\begin{equation}
T_{0}^{0}=\frac{b}{2m}\mathbf{\nabla }\Psi _{b}^{\ast }\mathbf{\nabla }\Psi
_{b}+\frac{\hbar ^{2}-b^{2}}{8m\rho ^{2}}\left( \mathbf{\nabla }\rho \right)
^{2}  \label{a1.14}
\end{equation}
\begin{equation}
T_{\alpha }^{0}=\frac{ib}{2}\left( \Psi _{b}^{\ast }\partial _{\alpha }\Psi
_{b}-\partial _{\alpha }^{\ast }\Psi _{b}\cdot \Psi _{b}\right) =mj_{\alpha
}=-mj^{\alpha },\qquad \alpha =1,2,3  \label{a1.15}
\end{equation}
Expressions for other components of the energy-momentum tensor are rather
complicated, and we do not write them down.

The action (\ref{a1.10}) describes the same dynamic system $\mathcal{S}_{%
\mathrm{q}}$ as the action (\ref{a1.2}) at any value of the constant $b\neq
0 $, because it is obtained from the action (\ref{a1.2}) by means of a
change of variables. Now dynamic equations (\ref{a1.11}) are nonlinear in
terms of the function $\Psi _{b}$ for all $b$ except for the case when $%
b^{2}=\hbar ^{2}$. Hence, dynamic system (\ref{a1.10}) ceases to be quantum,
if $b\neq \pm \hbar $, but the description of the quantum particle $\mathcal{%
S}_{\mathrm{q}}$ does not degenerate at $\hbar =0$. Elimination of the
degeneration at $\hbar =0$ is connected with the fact that the
transformation (\ref{a1.9}) depends on the parameter $\hbar $ analytically
for all values $\hbar $ except for $\hbar =0$.

Dependence of the function $\Psi _{b}$ on the arbitrary constant $b$ is
connected with the fact that the function $\Psi _{b}$ is some kind of
complex potential. A characteristic property of a potential is an expression
of physical quantities via derivatives of the potential. As a result any
physical state may be described by different potentials. The function $\Psi
_{b}$ is a potential in the sense that the same physical state of the
dynamic system $\mathcal{S}_{\mathrm{q}}$ may be described by different
functions $\Psi _{b}$. The electromagnetic potentials $A_{k}$, $k=0,1,2,3$
have the same property, because the state $\mathbf{E}$, $\mathbf{H}$ of the
electromagnetic field determines the electromagnetic potential within the
gauge transformation. Corresponding gauge transformation for the function $%
\Psi _{b}$ exists also, but it is more complicated, than that for
electromagnetic potentials \cite{R99}. In the given case the constant $b$ is
one of parameters of the gauge transformation for the function $\Psi _{b}$.
The gauge transformation connected with the change $b\rightarrow \tilde{b}$
has the form 
\begin{equation}
\Psi _{b}\rightarrow \Psi _{\tilde{b}}=\left| \Psi _{b}\right| \exp \left( 
\frac{b}{\tilde{b}}\log \frac{\Psi _{b}}{\left| \Psi _{b}\right| }\right)
\label{a1.15a}
\end{equation}

The action (\ref{a1.2}) is written in such a gauge, where $b=\hbar $. In
this case $\Psi _{b=\hbar }=\psi $, the dynamic equation is linear, the
quantum principles are fulfilled and the dynamic system is quantum. But this
gauge is unsuccessful from the viewpoint of transition to the classical
limit. Any other gauge $b\neq \hbar $ is successful from viewpoint of
transition to the classical limit, but it is unsuccessful from viewpoint of
quantum principles, because the dynamic equation (\ref{a1.11}) is nonlinear
in this case. In the gauge $b\neq \hbar $ the function $\Psi _{b}$ is not a
wave function in the sense, that it is not a vector of the Hilbert space. In
this case the function $\Psi _{b}$ is simply a complex dependent variable,
describing the dynamic system $\mathcal{S}_{\mathrm{q}}$.

Setting $\hbar =0$ in relations (\ref{a1.10}) - (\ref{a1.15}), we obtain the
classical approximation of the quantum particle $\mathcal{S}_{\mathrm{q}}$
description. We obtain the action for the dynamic system $\mathcal{C}$,
defined by (\ref{a2.1}) 
\begin{equation}
\mathcal{A}\left[ \Psi _{b},\Psi _{b}^{\ast }\right] =\int \left\{ \frac{ib}{%
2}\left( \Psi _{b}^{\ast }\partial _{0}\Psi _{b}-\partial _{0}\Psi
_{b}^{\ast }\cdot \Psi _{b}\right) -\frac{b^{2}}{2m}\mathbf{\nabla }\Psi
_{b}^{\ast }\mathbf{\nabla }\Psi _{b}+\frac{b^{2}}{8m}\left( \mathbf{\nabla }%
\left| \Psi _{b}\right| \right) ^{2}\right\} dtd\mathbf{x}  \label{a1.16}
\end{equation}
The dynamic system (\ref{a1.16}) may be considered as a limit of $\mathcal{S}%
_{\mathrm{q}}$ at $\hbar =0$. Of course, it is not a quantum system.
Nevertheless, if we describe the dynamic system $\mathcal{S}_{\mathrm{q}}$
in terms of the functions $\Psi _{b}$ with some fixed $b\neq 0$, we may add
the limiting system $\mathcal{C}=Q_{\hbar =0}$ to the set $\mathcal{SQ}$. As
a result we obtain the set $\mathcal{SG=SD\cup }\left\{ \mathcal{C}\right\} $%
, containing dynamic systems $\mathcal{S}_{\mathrm{q}}=Q_{\hbar }$ with all
values of the parameter $\hbar $, including $\hbar =0$.

Already at this stage of our investigation we can conclude, that the action (%
\ref{a1.16}) cannot describe an individual classical particle $\mathcal{S}_{%
\mathrm{cl}}$, because the action (\ref{a1.16}) describes a continuous
dynamic system, whereas $\mathcal{S}_{\mathrm{cl}}$ is a discrete system. In
other words, the transition $\hbar \rightarrow 0$ does not suppress degrees
of freedom of the dynamic system $\mathcal{S}_{\mathrm{q}}$. This result may
be formulated in the form.

\textit{The statement that the wave function describes the state of
individual particle (system) is incompatible with the quantum mechanics
formalism. }

This important conclusion is a rough one in the sense, that is does not
depend on whether or not the actions (\ref{a1.16}) and (\ref{a1.22})
describe the same dynamic system.

It was shown \cite{R99}, that the wave function (the function $\psi $) is
the method of description of any ideal fluid (or a fluidlike dynamic
system), but not a specific quantum object. Making change of variables and
using technique developed in \cite{R99}, one can reduce the action (\ref
{a1.22}) to a description in terms of the function $\psi $ (wave function).
Then it appears that the action (\ref{a1.16}) describes irrotational flows
of the fluid. In other words, the dynamic system $\mathcal{C}$ is a partial
case of the dynamic system $\mathcal{E}\left[ \mathcal{S}_{\mathrm{cl}}%
\right] $. In general case the arbitrary flow is described by the action (%
\ref{a1.22}) (See mathematical proof in Appendix)

Note, that the transformation (\ref{a1.9}) of the wave function phase is
well known (see, for instance, \cite{LL89}, Section 17). One has shown that
after the transformation (\ref{a1.9}) the quantum description turns to the
classical one, provided $\hbar \rightarrow 0$. However, the question what
object (an individual classical particle, or a statistical ensemble of
classical particles) is described by the obtained Hamilton-Jacobi equation
was not considered.

\section{Statistical interpretation or statistical ground?}

Thus, we have shown that the relation (\ref{a1.1b}) is incompatible with the
quantum mechanics formalism, whereas the relation (\ref{a1.1a}) is
fulfilled. The following question arises. What relation does take place
between the quantum stochastic particle $\mathcal{S}_{\mathrm{st}}$ and the
classical particle $\mathcal{S}_{\mathrm{cl}}$? May we set 
\begin{equation}
\mathcal{S}_{\mathrm{q}}=\mathcal{E}\left[ \mathcal{S}_{\mathrm{st}}\right]
,\qquad \text{at }\hbar \rightarrow 0,\;\;\;\mathcal{S}_{\mathrm{st}%
}\rightarrow \mathcal{S}_{\mathrm{cl}}?  \label{a3.1}
\end{equation}
May we consider that the relation (\ref{a1.1a}) is a corollary of (\ref{a3.1}%
)? In reality we cannot test the second relation (\ref{a3.1}), because $%
\mathcal{S}_{\mathrm{st}}$ is not a dynamic system, and we have no
information on $\mathcal{S}_{\mathrm{st}}$. As to the first relation (\ref
{a3.1}), we can only say that it is compatible with the action (\ref{a1.2}),
as far as the action (\ref{a1.2}) possesses the main property of the
statistical ensemble action $\mathcal{A}_{\mathcal{E}}$ 
\begin{equation}
\mathcal{A}_{\mathcal{E}}\left[ a\rho ,...\right] =a\mathcal{A}_{\mathcal{E}}%
\left[ \rho ,...\right] ,\qquad a=\text{const}>0  \label{a3.2}
\end{equation}
where $\rho $ is the density of elements (particles) in the statistical
ensemble. In the case of the action (\ref{a1.2}) $\rho =\psi ^{\ast }\psi $,
and the action (\ref{a1.2}) may be considered to be the action of the
statistical ensemble. The property (\ref{a3.2}) describes the fact that
elements (particles) of the statistical ensemble are independent. The action
of any quantum system has the property (\ref{a3.2}), because it is always
bilinear with respect to the wave function $\psi $. The property (\ref{a3.2}%
) is not violated at any change of variables. For instance, the action (\ref
{a1.10}) has the property (\ref{a3.2}), although the action (\ref{a1.10})
generates nonlinear dynamic equation (\ref{a1.11}) and, hence, it is
incompatible with the QM principles. On the other hand, the action (\ref
{a1.10}) describes the same dynamic system as the action (\ref{a1.2}), which
describes a quantum system compatible with the QM principles.

In fact, any interpretation of QM is connected closely with the QM
formalism. The statistical interpretation of quantum mechanics is
conditioned by the special formalism which distinguishes from the orthodox
QM technique. This formalism, generating the statistical interpretation, is
the statistical ensemble technique (SET). There is some distinction between
SET and the orthodox QM formalism. SET is a more general formalism, than the
QM technique. SET does not use QM principles. Introducing QM principles in
SET, we constraint SET, and it turns into the orthodox QM formalism. In the
conventional QM formalism the wave function is a specific quantum object, it
is a vector in the Hilbert space, whereas in SET the function $\psi $ (wave
function) is a method of the continuous dynamic system description. In
particular, a description in terms of the wave function is a method of the
pure statistical ensemble description.

The pure statistical ensemble of free classical (deterministic) particles is
a dynamic system $\mathcal{E}\left[ \mathcal{S}_{\mathrm{cl}}\right] $,
whose action has the form (\ref{a1.22}). If the particles are stochastic the
action (\ref{a1.22}) transforms to the form 
\begin{equation}
\mathcal{E}\left[ \mathcal{S}_{\mathrm{st}}\right] :\qquad \mathcal{A}_{%
\mathcal{E}\left[ \mathcal{S}_{\mathrm{st}}\right] }\left[ \mathbf{x,u}%
\right] =\int \left\{ \frac{m}{2}\left( \frac{d\mathbf{x}}{dt}\right) ^{2}+%
\frac{m}{2}\mathbf{u}^{2}-\frac{\hbar }{2}\mathbf{\nabla u}\right\} dtd%
\mathbf{\xi }  \label{b3.1}
\end{equation}
where $\mathbf{u}=\mathbf{u}\left( t,\mathbf{x}\right) $ is a vector
function of arguments $t,\mathbf{x}$ (not of $t,\mathbf{\xi }$), and $%
\mathbf{x}=\mathbf{x}\left( t,\mathbf{\xi }\right) $ is a vector function of
independent variables $t,\mathbf{\xi }$. The 3-vector $\mathbf{u}$ describes
the mean value of the stochastic component of the particle motion, which is
a function of the variables $t,\mathbf{x}$. The first term $\frac{m}{2}%
\left( \frac{d\mathbf{x}}{dt}\right) ^{2}$ describes the energy of the
regular component of the stochastic particle motion. The second term $m%
\mathbf{u}^{2}/2$ describes the energy of the random component of velocity.
The components $\frac{d\mathbf{x}}{dt}$ and $\mathbf{u}$ of the total
velocity are connected with different degrees of freedom, and their energies
should be added in the expression for the Lagrange function density. The
last term $-\hbar \mathbf{\nabla u}/2$ describes interaction between the
regular component $\frac{d\mathbf{x}}{dt}$ and random one $\mathbf{u}$. Note
that $m\mathbf{u}^{2}/2$ is a function of $t,\mathbf{x}$\textbf{. }It
influences on the regular component $\frac{d\mathbf{x}}{dt}$ as a potential
energy $U\left( t,\mathbf{x},\nabla \mathbf{x}\right) =-m\mathbf{u}^{2}/2$,
generated by the random component.

The dynamic system (\ref{b3.1}) is a statistical ensemble, because the
Lagrange function density of the action (\ref{b3.1}) does not depend on $%
\mathbf{\xi }$ explicitly, and we can represent the action for the single
system $\mathcal{S}_{\mathrm{st}}$%
\begin{equation}
\mathcal{S}_{\mathrm{st}}:\qquad \mathcal{A}_{\mathcal{S}_{\mathrm{st}}}%
\left[ \mathbf{x},\mathbf{u}\right] =\int \left\{ \frac{m}{2}\left( \frac{d%
\mathbf{x}}{dt}\right) ^{2}+\frac{m}{2}\mathbf{u}^{2}-\frac{\hbar }{2}%
\mathbf{\nabla u}\right\} dt  \label{b3.1c}
\end{equation}
Unfortunately, the expression for the action (\ref{b3.1c}) is only symbolic,
because the differential operator $\mathbf{\nabla }=\left\{ \partial
/\partial x^{\alpha }\right\} $, $\alpha =1,2,3$ is defined in the
continuous vicinity of the point $\mathbf{x}$, but not only for one point $%
\mathbf{x}$. The expression (\ref{b3.1c}) ceases to be symbolic, only if $%
\hbar =0$. In this case the last term, containing $\mathbf{\nabla }$
vanishes. Variation of (\ref{b3.1c}) with respect to $\mathbf{u}$ gives $%
\mathbf{u}=0$, and the action (\ref{b3.1c}) coincides with the action (\ref
{a1.21}) for $\mathcal{S}_{\mathrm{cl}}$. If $\hbar \neq 0$, the expression
for the action (\ref{b3.1c}) is not the well defined, and dynamic equations
for $\mathcal{S}_{\mathrm{st}}$ are absent.

If the quantum constant $\hbar =0$, it follows from the dynamic equation for 
$\mathbf{u}$, that $\mathbf{u}=0$, and the action (\ref{b3.1}) reduces to
the form (\ref{a1.22}). The dynamic system $\mathcal{S}_{\mathrm{q}}$ with
the action (\ref{a1.2}) is a special case of the dynamic system (\ref{b3.1}%
). To obtain this result we should use SET, described in \cite{R99,R2002}.
Application of SET to the statistical ensemble $\mathcal{E}\left[ \mathcal{S}%
_{\mathrm{cl}}\right] $ can be found in Appendix. In our discussion we shall
not go into mathematical detail and restrict ourselves by the conceptual
consideration.

Having written the action (\ref{b3.1}), we set the problem of the stochastic
particle motion mathematically. Interpretation of the obtained solution for
dynamic variables $\mathbf{x}\left( t,\mathbf{\xi }\right) $ and $\mathbf{u}%
\left( t\mathbf{,x}\right) $ is also clear. The quantity $\mathbf{x}\left( t,%
\mathbf{\xi }\right) $ describes the mean trajectories of the ensemble
particles and motion along them. We may say that $\mathbf{x}\left( t,\mathbf{%
\xi }\right) $ describes the motion of the statistical average particle $%
\left\langle \mathcal{S}_{\mathrm{st}}\right\rangle $. Constructing the
canonical energy-momentum tensor, we can attribute the mean momentum and the
mean energy to the statistical average particle $\left\langle \mathcal{S}_{%
\mathrm{st}}\right\rangle $. We can describe only mean motion of particles.
Real motion of the stochastic particle cannot be described by the
consideration of dynamic equation for $\mathcal{E}\left[ \mathcal{S}_{%
\mathrm{st}}\right] $, as well as we cannot describe the real motion of the
gas molecules in the framework of the gas dynamic equations (Euler
equations). We can describe only the mean motion, i.e. the motion of the
''gas particles'', containing many molecules.

The action (\ref{b3.1}) does not contain time derivatives of $\mathbf{u}$.
It means that $\mathbf{u}$ is a function of the ensemble state, and
evolution of $\mathbf{u}$ is determined by the evolution of the variables $%
\mathbf{x}$. The variable $\mathbf{u}$ can be eliminated easily. Dynamic
equations for $\mathbf{u}$ are obtained by variation of the action (\ref
{b3.1}) with respect to $\mathbf{u}$. As far as $\mathbf{u}$ is a function
of $t,\mathbf{x}$, before the variation one should transform the action (\ref
{b3.1}) to independent variables $t,\mathbf{x}$. After such a transformation
and variation with respect to $\mathbf{u}$ we obtain

\begin{equation}
\frac{\delta \mathcal{A}_{\mathcal{E}\left[ \mathcal{S}_{\mathrm{stl}}\right]
}}{\delta \mathbf{u}}=m\mathbf{u}\rho +\frac{\hbar }{2}\mathbf{\mathbf{%
\nabla }}\rho =0,  \label{b3.1a}
\end{equation}
where 
\begin{equation}
\rho =\rho \left( t,\mathbf{x}\right) =\frac{\partial \left( \xi _{1},\xi
_{2},\xi _{3}\right) }{\partial \left( x^{1},x^{2},x^{3}\right) }=\left( 
\frac{\partial \left( x^{1},x^{2},x^{3}\right) }{\partial \left( \xi
_{1},\xi _{2},\xi _{3}\right) }\right) ^{-1}  \label{b3.1b}
\end{equation}
Here $\mathbf{x}=\mathbf{x}\left( t,\mathbf{\xi }\right) $. The density $%
\rho =\rho \left( t,\mathbf{x}\right) $ is a complicated function of
derivatives of $\mathbf{x}$ with respect to $\mathbf{\xi }$, expressed as a
function of $t,\mathbf{x}$. Dynamic equations (\ref{b3.1a}) can be resolved
with respect to $\mathbf{u}$%
\begin{equation}
\mathbf{u}=\mathbf{u}\left( t,\mathbf{x}\right) =-\frac{\hbar }{2m}\mathbf{%
\nabla }\ln \rho \left( t,\mathbf{x}\right) ,  \label{b3.2}
\end{equation}
Using the solution (\ref{b3.2}), we can eliminate the variable $\mathbf{u}$ 
\cite{R2003}. Instead of (\ref{b3.1}) we obtain 
\begin{equation}
\mathcal{E}\left[ \mathcal{S}_{\mathrm{st}}\right] :\qquad \mathcal{A}_{%
\mathcal{E}\left[ \mathcal{S}_{\mathrm{st}}\right] }\left[ \mathbf{x}\right]
=\int \left\{ \frac{m}{2}\left( \frac{d\mathbf{x}}{dt}\right) ^{2}-U\left(
\rho ,\mathbf{\nabla }\rho \right) \right\} dtd\mathbf{\xi }  \label{b3.3}
\end{equation}
\begin{equation}
U\left( \rho ,\mathbf{\nabla }\rho \right) =\frac{\hbar ^{2}}{8m}\left( 
\mathbf{\nabla }\ln \rho \right) ^{2},\qquad \rho =\rho \left( t,\mathbf{x}%
\right)  \label{b3.4}
\end{equation}
where $\mathbf{x}=\mathbf{x}\left( t,\mathbf{\xi }\right) $ and $\rho =\rho
\left( t,\mathbf{x}\right) $ is determined by the last relation (\ref{b3.1b}%
).

Using SET \cite{R99,R2002}, we can transform the action (\ref{b3.3}) to
dependent variables $\psi ,\psi ^{\ast }$%
\begin{eqnarray}
\mathcal{A}_{\mathcal{E}\left[ \mathcal{S}_{\mathrm{st}}\right] }\left[ \psi
,\psi ^{\ast }\right] &=&\int \left\{ \frac{ib}{2}\left( \psi ^{\ast
}\partial _{0}\psi -\partial _{0}\psi ^{\ast }\cdot \psi \right) -\frac{b^{2}%
}{2m}\mathbf{\nabla }\psi ^{\ast }\mathbf{\nabla }\psi \right.  \nonumber \\
&&\left. +\frac{b^{2}}{8m}\sum\limits_{\alpha =1}^{\alpha =3}(\mathbf{\nabla 
}s_{\alpha })^{2}\rho +\frac{b^{2}-\hbar ^{2}}{8\rho m}(\mathbf{\nabla }\rho
)^{2}\right\} dtd\mathbf{x},  \label{b3.5}
\end{eqnarray}
where 
\begin{equation}
\psi =\left\{ \psi _{1},\psi _{2}\right\} ,\qquad \psi ^{\ast }=\left\{ 
\begin{array}{c}
\psi _{1}^{\ast } \\ 
\psi _{2}^{\ast }
\end{array}
\right\} ,\qquad \rho \equiv \psi ^{\ast }\psi ,\qquad s_{\alpha }\equiv 
\frac{\psi ^{\ast }\sigma _{\alpha }\psi }{\rho },\qquad \alpha =1,2,3,
\label{b3.6}
\end{equation}
and $\sigma _{\alpha }$ are the Pauli matrices.

If the components of the function $\psi $ are linear dependent: $a_{1}\psi
_{1}+a_{2}\psi _{2}=0,$\ \ $a_{1},a_{2}=$const, the quantities $s_{\alpha }=$%
const, $\alpha =1,2,3$, and the third term in (\ref{b3.5}) vanishes. In this
special (irrotational) case the action coincides with the action (\ref{a1.10}%
) for $\mathcal{S}_{\mathrm{q}}$.

The action (\ref{a1.10}) is a special case of the statistical ensemble of
stochastic (and classical, if $\hbar =0$) particles. A direct interpretation
of the action (\ref{a1.10}) is absent. It should be interpreted, using the
fact that the dynamic system (\ref{a1.10}) is a special case of the dynamic
system (\ref{b3.1}). The dynamic equation (\ref{a1.11}), generated by the
action (\ref{a1.10}) is nonlinear, and the dynamic system (\ref{a1.10}) is
not quantum in the sense, that it is incompatible with the QM principles.

From the pure mathematical viewpoint the situation looks as follows. The
action (\ref{b3.3}) describes the statistical ensemble, because the
functional (\ref{b3.3}) has the property (\ref{a3.2}), which is the main
property of the statistical ensemble. Besides, it is equivalent to the
action (\ref{b3.1}) for the statistical ensemble. Dynamic equations,
generated by the action (\ref{b3.3}) are the partial differential equations.
If $\hbar =0$, these equations can be reduced to a system of ordinary
differential equations, as one can see directly from (\ref{b3.3}). In the
general case $\hbar \neq 0$ such a reduction is impossible. This fact is
interpreted in the sense, that in the special case $\hbar =0$ the
statistical ensemble (\ref{b3.3}) turns to the statistical ensemble of
classical particles.

On the one hand, the functional (\ref{b3.3}) is the action for the
statistical ensemble $\mathcal{E}\left[ \mathcal{S}_{\mathrm{st}}\right] $.
On the other hand, the functional (\ref{b3.3}) is the action for some
continuous set $\mathcal{S}_{\mathrm{eff}}\left[ \mathcal{S}_{\mathrm{cl}}%
\right] $ of classical particles $\mathcal{S}_{\mathrm{cl}}$, labelled by
the parameter $\mathbf{\xi }$. This set $\mathcal{S}_{\mathrm{eff}}\left[ 
\mathcal{S}_{\mathrm{cl}}\right] $ is not a statistical ensemble of
classical particles $\mathcal{S}_{\mathrm{cl}}$, because these classical
particles interact between themselves by means of the potential energy $%
U\left( \rho ,\mathbf{\nabla }\rho \right) $, determined by the relation (%
\ref{b3.4}) and, hence, these classical particles are not independent. If $%
\hbar =0$, this interaction vanishes, the set $\mathcal{S}_{\mathrm{eff}}%
\left[ \mathcal{S}_{\mathrm{cl}}\right] \,\,$of classical particles $%
\mathcal{S}_{\mathrm{cl}}$ turns to the statistical ensemble $\mathcal{E}%
\left[ \mathcal{S}_{\mathrm{cl}}\right] $ of classical particles $\mathcal{S}%
_{\mathrm{cl}}$, because the particles $\mathcal{S}_{\mathrm{cl}}$ become to
be independent.

SET describes the long chain of the dynamic variables transformation, which
leads from the action (\ref{b3.1}) for the statistical ensemble $\mathcal{E}%
\left[ \mathcal{S}_{\mathrm{st}}\right] $ of stochastic particles $\mathcal{S%
}_{\mathrm{st}}$ to the action (\ref{a1.2}) for the quantum particle $%
\mathcal{S}_{\mathrm{q}}$. This chain contains integration of the system of
dynamic equations (\ref{A.8}). The integration is rather complicated (at
least, it was unknown for a long time). The dynamic equations (\ref{A.8}) do
not depend on the form of the Lagrangian of the classical system $\mathcal{S}%
_{\mathrm{cl}}$, and they can be integrated for any Lagrangian $L\left( t,%
\mathbf{x}\right) $, but not only for $L\left( t,\mathbf{x}\right) =m\left( 
\frac{d\mathbf{x}}{dt}\right) ^{2}/2$, as we have done in the present paper.

Note that the statistical interpretation of quantum mechanics is based on
the fact that the dynamic system $\mathcal{S}_{\mathrm{q}}$ (\ref{a1.2}) is
a partial case of the statistical ensemble $\mathcal{E}\left[ \mathcal{S}_{%
\mathrm{st}}\right] $, described by the action (\ref{b3.1}). For such an
interpretation we \textit{need neither QM principles, nor other additional
suppositions}.

The representation (\ref{b3.1}) is explicitly statistical, and the
statistical interpretation of quantum mechanics may be obtained from the
actions (\ref{b3.1}), or (\ref{b3.3}). But these actions generate dynamic
equations, which are difficult for solution and for investigation. The
beginning of the chain is easy for interpretation, but it is difficult for
investigation. On the contrary, the end of the chain (the action (\ref{a1.2}%
) for $\mathcal{S}_{\mathrm{q}}$) is easy for solution, and it is difficult
for interpretation, as far as the action (\ref{a1.2}) does not remind the
action for the statistical ensemble. Thus, for solution of dynamic equation
and for a correct interpretation of these solutions we need both actions (%
\ref{a1.2}), (\ref{b3.1}) and mathematical relations between them.
Interpretation of quantum mechanics obtained on the basis of verbal
considerations and reasonings \cite{P64}-\cite{G90} cannot compete with the
statistical interpretation, obtained on the basis of the chain of exact
mathematical relations. For instance, even the most developed Bohmian
hydrodynamical interpretation \cite{B52} of QM uses only a part of the chain
of relations between the actions (\ref{b3.1}) and (\ref{a1.2}). Besides, in
the Bohmian interpretation one can move along the chain only in one
direction: from the action (\ref{a1.2}) to the action (\ref{b3.1}), i.e.
from the wave function to hydrodynamics. The motion along the chain in the
opposite direction was blocked, because the result of integration (\ref{A.11}%
) of dynamic equations (\ref{A.8}) was known only for the case of
irrotational flows. As a result for the Bohmian interpretation of QM uses
the QM principles, reformulated in terms of hydrodynamic variables.

It is very difficult to produce a sequential statistical interpretation of
the quantum mechanics on the verbal level without a use of exact
mathematical relations of the chain (\ref{b3.1}) -- (\ref{a1.2}). It is easy
to make mistakes, especially, if we do not take into account that the
statistical description of quantum mechanics is a \textit{non-probabilistic
statistical description}. In the statistical physics we meet only
statistical description in terms of the probability theory, and some
researchers believe that terms ''statistical'' and ''probabilistic'' are
synonyms. In reality, the term ''statistical description'' means the
description, dealing with many similar, or almost similar objects, whereas
the term ''probabilistic description'' means a logical construction, founded
on the probability theory.

In the probability theory the probability and the number of objects must be
necessarily nonnegative. This constraint can be fulfilled, if the objects of
the statistics can be represented as points in some space (for instance, the
phase space in the statistical physics). But if the objects of the
statistics are extended objects, for instance, world lines, their density
(and number) can be negative. Indeed, the density $j^{k}$ of world lines in
the vicinity of the space-time point $x$ is determined by the relation 
\begin{equation}
dN=j^{k}dS_{k}  \label{b3.7}
\end{equation}
where $dN$ is the flux of the world lines through the infinitesimal area $%
dS_{k}$, and $j^{k}$ is the proportionality coefficient between the two
quantities. The quantity $j^{k}=j^{k}\left( x\right) $ is by definition the
world lines density in the vicinity of the point $x$. The quantity $j^{k}$
can be negative, and the probabilistic description of world lines appears to
be impossible. On the other hand, the statistical description in terms of
world lines cannot be replaced by the statistical description in terms of
particles, because the last description is nonrelativistic.

In the relativity theory the world line of a particle is a real physical
object, whereas the particle is an attribute of the world line (intersection
of the world line with the plane $t=$const). In the nonrelativistic theory,
where the absolute simultaneity is supposed to exist, the particle may be
considered to be a real physical object, and world line is an attribute of
the particle (its history). Stochastic component of motion is relativistic
even in the nonrelativistic QM, and we should use the relativistic
statistical description even in the nonrelativistic QM, where only the mean
particle motion (regular component of motion) is nonrelativistic. In this
sense the nonrelativistic QM has relativistic roots.

In general, the probabilistic statistical interpretation is more
informative, than the non-probabilistic one, because, using the probability
theory, one can obtain such distributions, which cannot be obtained in the
non-probabilistic statistical description. In the quantum mechanics the
statistical description is obtained by description and investigation of the
statistical ensemble, which is investigated simply as a continuous dynamic
system. To realize this program, it is very useful to introduce the concept
of the statistical average object (particle). Mathematically it is carried
out as follows.

The number $N$ of particles $\mathcal{S}$, constituting the statistical
ensemble $\mathcal{E}\left[ N,\mathcal{S}\right] $, is supposed to be large
enough, and the properties of the statistical ensemble do not depend on the
number $N$ of its elements $\mathcal{S}$ (the property (\ref{a3.2}) of the
statistical ensemble). As far as the properties of the statistical ensemble
do not depend on $N$, we can set formally $N=1$ and introduce the concept of
the statistical average particle $\left\langle \mathcal{S}\right\rangle =%
\mathcal{E}\left[ 1,\mathcal{S}\right] $, which is by definition the
statistical ensemble normalized to one particle. Although the number $N$ of
particles in $\left\langle \mathcal{S}\right\rangle =\mathcal{E}\left[ 1,%
\mathcal{S}\right] $ is equal to $1$, the statistical average particle $%
\left\langle \mathcal{S}\right\rangle $ is the statistical ensemble, and $%
\left\langle \mathcal{S}\right\rangle $ has properties of the statistical
ensemble. In particular, $\left\langle \mathcal{S}\right\rangle $ has
infinite number of degrees of freedom.

The statistical ensemble $\mathcal{E}\left[ N,\mathcal{S}\right] $ ($%
N\rightarrow \infty $) of $N$ particles $\mathcal{S}$ is the dynamic system,
whose action is $\mathcal{A}_{\mathcal{E}\left[ N,\mathcal{S}\right] }$. The
statistical average particle $\left\langle \mathcal{S}\right\rangle $ is the
dynamic system, whose action $\mathcal{A}_{\left\langle \mathcal{S}%
\right\rangle }$. The actions $\mathcal{A}_{\left\langle \mathcal{S}%
\right\rangle }$ and $\mathcal{A}_{\mathcal{E}\left[ N,\mathcal{S}\right] }$
are connected by the relation 
\begin{equation}
\mathcal{A}_{\left\langle \mathcal{S}\right\rangle }=\lim_{N\rightarrow
\infty }\frac{1}{N}\mathcal{A}_{\mathcal{E}\left[ N,\mathcal{S}\right] }
\label{a1.1}
\end{equation}
Thus, we can speak about the statistical average particle $\left\langle 
\mathcal{S}\right\rangle $ instead of the statistical ensemble $\mathcal{E}%
\left[ \mathcal{S}\right] $.

At the same time because of normalization to one particle, the statistical
average particle $\left\langle \mathcal{S}\right\rangle $ may be perceived
as a diffuse individual particle. The energy-momentum vector $P_{k}$ of $%
\left\langle \mathcal{S}\right\rangle $, considered as a dynamic system, may
be associated with the energy-momentum $P_{k}$ of the individual particle $%
\mathcal{S}$. The same concerns the angular momentum and other additive
quantities.

If we solve dynamic equations for the statistical ensemble (\ref{b3.3}), we
obtain 
\begin{equation}
\mathbf{x}=\mathbf{x}\left( t,\mathbf{\xi }\right)  \label{b3.8}
\end{equation}
At fixed $\mathbf{\xi }$ the relation (\ref{b3.8}) describes some world
line, which may be interpreted as a mean world line of a stochastic particle 
$\mathcal{S}$, or as a world line of the statistical average particle $%
\left\langle \mathcal{S}\right\rangle $. Such interpretation is possible,
because the statistical ensemble $\mathcal{E}\left[ N,\mathcal{S}\right] $, $%
N\rightarrow \infty $ may be considered: (1) as a statistical ensemble of $N$
stochastic particles $\mathcal{S}$ and (2) as a statistical ensemble of $N$
statistical average particles (dynamic systems) $\left\langle \mathcal{S}%
\right\rangle $. In other words,

\begin{equation}
\mathcal{E}\left[ N,\mathcal{S}\right] =\mathcal{E}\left[ N,\left\langle 
\mathcal{S}\right\rangle \right]  \label{b3.9}
\end{equation}

Thus, the statistical description of the quantum mechanics allows one to
obtain the mean world lines and to attribute to them some mean
energy-momentum. In the same way in the gas dynamics we can attribute the
world lines and the energy-momentum to the gas particles, which contains
many molecules. But in QM we cannot attribute the momentum distribution to
the stochastic particles, as well as in the framework of the gas dynamics we
cannot attribute the Maxwell distribution to the gas molecules, because the
gas dynamics formalism does not allow one to do this. The orthodox
interpretation of the quantum mechanics claims that it can predict the
momentum distribution of the particles on the basis of the QM principles,
but this claim is unwarranted, because this momentum distribution is
fictitious. This momentum distribution cannot be attributed to any definite
state (to a definite wave function), because in the experimental
measuremement of the momentum distribution the measurement of a single
momentum needs a long time. The state (wave function) changes essentially in
this time. (See detailed discussion in \cite{R2004}). In reality, the
orthodox QM formalism founded on the QM principles cannot give more, than
the statistical description, based on the SET, can give, because from the
mathematical viewpoint the QM principles are only a series of constraints.
imposed on possible dependent variables. For instance, the actions (\ref
{a1.2}) and (\ref{a1.10}) differ only in their dependent variables. In the
same time (\ref{a1.2}) satisfies the QM principles, whereas (\ref{a1.10})
does not.

The statistical average object may have properties which are alternative for
individual objects, and this property is a statistical property, but not a
special quantum property. For instance, the statistical average habitant of
a country is a hermaphrodite (half-man -- half-woman), whereas any
individual habitant of the same country is either a man, or a woman. If the
quantum particle $\mathcal{S}_{\mathrm{q}}$ is the statistical average
particle $\left\langle \mathcal{S}\right\rangle $, it can pass through two
slits simultaneously, whereas the individual particle $\mathcal{S}$ can pass
only through one of slits, and there is nothing mystic in this fact. If the
Schr\"{o}dinger cat is the statistical average $\left\langle \text{cat}%
\right\rangle $, it may be dead and alive simultaneously, although the
individual cat may be either alive, or dead.

If the intensity of the particle beam in the two-slit experiment is very
low, only one real particle appears in the space between the slits and the
screen. In this case the property of a statistical average particle to pass
through two slits simultaneously seems to be rather evident statistically,
but it cannot be explained from the viewpoint of the probability theory,
because one cannot introduce probability in the proper way. Sometimes one
considers this fact as a defect of the statistical interpretation, but it is
not a defect, because, as we have mentioned above, the quantum mechanics is
a \textit{non-probabilistic statistical conception}.

\section{Rapprochement of the statistical interpretation and the Copenhagen
one}

Conventionally in the classical dynamics the discrete dynamic system $%
\mathcal{S}_{\mathrm{cl}}$ (individual particle) is considered to be the
principal object of dynamics. The discrete dynamic system $\mathcal{S}_{%
\mathrm{cl}}$ is described by the action 
\begin{equation}
\mathcal{S}_{\mathrm{cl}}:\qquad \mathcal{A}_{\mathcal{S}_{\mathrm{cl}}}%
\left[ \mathbf{x}\right] =\int L\left( \mathbf{x,}\frac{d\mathbf{x}}{dt}%
\right) dt  \label{a2.2}
\end{equation}
where $\mathbf{x=x}\left( t\right) $ and $L\left( \mathbf{x,}\frac{d\mathbf{x%
}}{dt}\right) $ is the Lagrange function of $\mathcal{S}_{\mathrm{cl}}$.

The pure statistical ensemble $\mathcal{E}\left[ \mathcal{S}_{\mathrm{cl}}%
\right] $, whose action has the form 
\begin{equation}
\mathcal{E}\left[ \mathcal{S}_{\mathrm{cl}}\right] :\qquad \mathcal{A}_{%
\mathcal{E}\left[ \mathcal{S}_{\mathrm{cl}}\right] }\left[ \mathbf{x}\right]
=\int L\left( \mathbf{x,}\frac{d\mathbf{x}}{dt}\right) dtd\mathbf{\xi }
\label{a2.4}
\end{equation}
where $\mathbf{x}=\mathbf{x}\left( t,\mathbf{\xi }\right) $ is considered to
be a derivative object of dynamics, because it consists of many principal
objects of dynamics $\mathcal{S}_{\mathrm{cl}}$. The statistical ensemble $%
\mathcal{E}\left[ \mathcal{S}_{\mathrm{cl}}\right] $ has infinite number of
the freedom degrees. The Lagrange function $L$ of $\mathcal{S}_{\mathrm{cl}}$
is the Lagrange function density of $\mathcal{E}\left[ \mathcal{S}_{\mathrm{%
cl}}\right] $, and there is one-to-one correspondence between the dynamic
systems (\ref{a2.2}) and (\ref{a2.4}). The action $\mathcal{A}_{\left\langle 
\mathcal{S}_{\mathrm{cl}}\right\rangle }$ of the statistical average
particle (dynamic system) $\left\langle \mathcal{S}_{\mathrm{cl}%
}\right\rangle $ is connected with the action $\mathcal{A}_{\mathcal{E}\left[
\mathcal{S}_{\mathrm{cl}}\right] }$ of the statistical ensemble $\mathcal{E}%
\left[ \mathcal{S}_{\mathrm{cl}}\right] $ by means of relation (\ref{a1.1}),
and there is one-to-one correspondence between $\left\langle \mathcal{S}_{%
\mathrm{cl}}\right\rangle $ and $\mathcal{S}_{\mathrm{cl}}$.

In such a situation the continuous dynamic system $\left\langle \mathcal{S}_{%
\mathrm{cl}}\right\rangle $ may be considered to be the principal object of
classical dynamics. Then the discrete dynamic system $\mathcal{S}_{\mathrm{cl%
}}$ may be regarded as a derivative object: the partial case of the
statistical average dynamic system $\left\langle \mathcal{S}_{\mathrm{cl}%
}\right\rangle $, when distributions of all quantities in $\left\langle 
\mathcal{S}_{\mathrm{cl}}\right\rangle $ are $\delta $-like.

Such a redefinition of the concept of the principal object in the classical
dynamics is useful in two aspects.

1. In the dynamics of stochastic systems $\mathcal{S}_{\mathrm{st}}$ the
principal object of dynamics is the statistical average system $\left\langle 
\mathcal{S}_{\mathrm{st}}\right\rangle $, because $\mathcal{S}_{\mathrm{st}}$
is not a dynamic system, and it cannot be an object of dynamics, whereas the
statistical average system $\left\langle \mathcal{S}_{\mathrm{st}%
}\right\rangle $ is always a continuous dynamic system. In the dynamics of
classical systems $\mathcal{S}_{\mathrm{cl}}$ both systems $\left\langle 
\mathcal{S}_{\mathrm{cl}}\right\rangle $ and $\mathcal{S}_{\mathrm{cl}}$ are
dynamic systems, but $\left\langle \mathcal{S}_{\mathrm{cl}}\right\rangle $
is always a continuous dynamic system, whereas $\mathcal{S}_{\mathrm{cl}}$
may be a discrete dynamic system. Defining in classical dynamics the
continuous dynamic system $\left\langle \mathcal{S}_{\mathrm{cl}%
}\right\rangle $ as the principal object of dynamics, we obtain the uniform
definition of the principal object in the classical dynamics and in the
dynamics of stochastic systems. This uniform definition allows one to
consider the classical dynamics as a partial case of the stochastic dynamics
(dynamics of stochastic systems), which appears, when the stochasticity
intensity vanishes. We may introduce the concept of the physical system $%
\mathcal{S=}\left\{ \mathcal{S}_{\mathrm{cl}},\mathcal{S}_{\mathrm{st}%
}\right\} $, which is a collective concept with respect to concepts of the
discrete dynamic system $\mathcal{S}_{\mathrm{cl}}$ and the stochastic
system $\mathcal{S}_{\mathrm{st}}$. The uniform definition of the principal
object of dynamics is connected closely with the mathematical formalism of
dynamics, which appears to be common for all physical systems $\mathcal{S=}%
\left\{ \mathcal{S}_{\mathrm{cl}},\mathcal{S}_{\mathrm{st}}\right\} $. This
formalism is the statistical ensemble technique (SET), and it is quite
reasonable, as far as the statistical average system $\left\langle \mathcal{S%
}\right\rangle $ (i.e. the statistical ensemble) is the principal object of
dynamics. As we have seen, SET is series of transformations of dynamic
equations (or actions) for the statistical ensemble, which include changes
of dynamic variables and integration of some dynamic equations.

The situation can be manifested in the example of the one-atom gas flow.
Motion of the gas molecules is random. It is described by the Maxwell
distribution. The gas dynamics equations do not deal with molecules. They
describe motion of the gas particles. Any gas particle consists of $N$
molecules ($N\rightarrow \infty $) and has formally $6N$ degrees of freedom,
but the gas dynamics ignores all these degrees of freedom except for those
six of them, which describe the motion of the gas particle as a whole.
Although the molecular motion is random, the motion of the gas particles is
deterministic (not random), and one may say, that the gas particle is the
''statistical average molecule''. One can see some analogy between the
molecule and the gas particle on the one side and the stochastic system $%
\mathcal{S}_{\mathrm{st}}$ and the statistical average system $\left\langle 
\mathcal{S}_{\mathrm{st}}\right\rangle $ on the other side. Equations of the
gas dynamics form a closed system of dynamic equations, as well as the
dynamic equations of the quantum mechanics. Both systems of equations
describe the mean motion of the gas particles (in the case of the gas
dynamics) and that of the statistical average particle (in the case of the
quantum mechanics). In the case of the gas motion we have also the kinetic
theory, which describes an evolution of the Maxwell distribution. In the
case of the quantum particle, such a detailed description is absent now.

2. In all statistical conceptions (in physics, biology, sociology, etc.) one
introduces two objects: (1) an individual (stochastic) object $\mathcal{S}$
and (2) the statistical average object $\left\langle \mathcal{S}%
\right\rangle $. In any statistical conception all laws are established for $%
\left\langle \mathcal{S}\right\rangle $, and all predictions on the ground
of these laws are made for $\left\langle \mathcal{S}\right\rangle $. The
term ''statistical average object'' describes a very important notion in any
statistical conception. Although the term ''normalized to unity statistical
ensemble of particles '' and the term ''statistical average particle'' mean
the same object, the second term is more convenient in the following sense.
The main word in the first term is ''ensemble'', whereas the main word in
the second term is ''particle''. If for brevity we use abbreviation of the
term, we retain the main word of the term, omitting attributives. Let us
consider, for instance, the expression ''the state of electron is described
by the wave function $\psi $''. Expression of such a type can be found
practically in any paper on quantum mechanics. If the concept of the
statistical average object is not introduced, this expression seems to be
meaningless (from viewpoint of the statistical interpretation), because the
wave function can describe only the state of statistical ensemble, but not
that of individual electron. But if the conception of the statistical
average object is introduced, the term ''electron'' may be interpreted as an
abbreviation of the term ''statistical average electron''. In this case the
considered expression has the meaning, which is compatible with the
statistical interpretation. The use of the term ''statistical average
object'' reduces the difference between the Copenhagen interpretation and
the statistical one. It allows one to use the physical jargon, which is
suited for advocates of both (Copenhagen and statistical) interpretations.
In other words, statement of the Copenhagen interpretation that ''the wave
function describes the state of the particle'' and other expressions of the
same sort becomes to be quite correct, provided the term ''particle'' means
some ''quantum particle'', but not the usual classical particle. This
''quantum particle'' is in reality the statistical average particle, having
the statistical properties of the ensemble.

In many cases of the gas dynamics we can speak about deterministic motion of
the gas particles, considering them as classical particles (of six freedom
degrees). In this case we ignore the fact that any gas particle is a
complicated object, consisting of many molecules and having many degrees of
freedom. In similar way, in the case of quantum mechanics we can ignore the
fact that the statistical average particle $\left\langle \mathcal{S}%
\right\rangle $ is a continuous dynamic system. Formally it is also
connected with the possible abbreviation of the term ''statistical average
particle'', when the main word of the term is retained.

\section{Discussion}

Thus, consideration of the statistical average object $\left\langle \mathcal{%
S}\right\rangle =\left\{ \left\langle \mathcal{S}_{\mathrm{cl}}\right\rangle
,\left\langle \mathcal{S}_{\mathrm{st}}\right\rangle \right\} $ as the
principal object of dynamics allows one to consider the quantum mechanics
and classical dynamics as two partial cases of the stochastic system
dynamics. The dynamics of stochastic systems is a statistical construction,
which starts from the action (\ref{b3.1}) for the pure statistical ensemble
of stochastic systems. Replacing Lagrange function for free particle by the
Lagrange function $L$ in the primary action (\ref{b3.1}), we obtain quantum
description for the physical system, whose classical description is
determined by this Lagrange function $L$. Replacing two last ''stochastic''
terms in the primary action (\ref{b3.1}), we obtain the statistical
description of the stochastic system with other type of stochasticity, than
the quantum stochasticity. Formally the type of stochasticity is described
by the form of the interaction term in (\ref{b3.4}) in the action (\ref{b3.3}%
) for the continuous set of classical interacting particles (systems).

The dynamic equation, generated by the actions (\ref{b3.1}), or (\ref{b3.3}%
), are difficult for solution. Choosing dynamic variables in a proper way,
we can obtain linear equations of the quantum mechanics, which are easier
for solution and investigation, than the primary dynamic equations,
generated by the actions (\ref{b3.1}), or (\ref{b3.3}). The QM principles
institutionalize how these variables should be chosen, to obtain linear
dynamic equations. The constraints of the QM principles are rigid enough to
obtain the correct nonrelativistic dynamic equations without a reference to
the primary statistical description. But the QM principles in themselves
cannot give a correct statistical interpretation. Besides, they cannot give
a correct extension of nonrelativistic QM into the relativistic region,
because the QM principles are nonrelativistic. Applying the orthodox QM to
the problem of the relativistic particle collision, we obtain only the $S$%
-matrix consideration, whereas a use of the primary actions (\ref{b3.1}), or
(\ref{b3.3}) admits one to obtain more detailed picture in the region of
collision \cite{R003}. In general, the QM principle of linearity of dynamic
equations is not a physical principle, because the principle of the logical
simplicity and the principle of simplicity of the dynamic equation solution
are quite different things.

As we have mentioned, in any statistical conception there are two sorts of
object: individual $\mathcal{S}$ and statistical average $\left\langle 
\mathcal{S}\right\rangle $. Formalism of the statistical description and of
the quantum mechanics admits one to consider only the statistical average
object $\left\langle \mathcal{S}\right\rangle $ and to make predictions
concerning $\left\langle \mathcal{S}\right\rangle $. The individual
stochastic object retains outside the framework of the formalism. But we
cannot ignore the individual stochastic system $\mathcal{S}$ completely,
because it appears in the measurements, and generates problems, connected
with the QM interpretation. The main problem of such a kind is the problem
of the wave function reduction at the measurement.

Identification $\mathcal{S}_{\mathrm{q}}=\left\langle S_{\mathrm{st}%
}\right\rangle $ eliminates all problems connected with the interpretation
of the wave function reduction. The correct interpretation is connected with
the correct interpretation of the concept of the measurement. Existence of
two different objects $\mathcal{S}$ and $\left\langle \mathcal{S}%
\right\rangle $ in any statistical conception generates two sorts of
measurements: $S$-measurement and $M$-measurement, which are produced
respectively under $\mathcal{S}$ and under $\left\langle \mathcal{S}%
\right\rangle $. The mathematical technique of QM deals only with the
statistical average particles $\left\langle S\right\rangle $, and all
predictions of QM have a probabilistic (statistical) character. The quantum
technique can predict distribution $F\left( R\right) $ of the quantity $%
\mathcal{R}$ at the state $\psi $, or the probability $w\left( R\right) $ of
the quantity $\mathcal{R}$ at the state $\psi $. Validity of the prediction
can be tested by the measurement. But this experiment is the massive
experiment ($M$-measurement), i.e. a set of many single measurements ($S$%
-measurements), because neither distribution $F\left( R\right) $, nor the
probability $w\left( R\right) $ can be measured by means of a single
measurement of the quantity $\mathcal{R}$. Thus, measurement in quantum
mechanics is the $M$-measurement, consisting of many $S$-measurements.

Even if at the state $\psi $ the quantity $\mathcal{R}$ has the unique value 
$R^{\prime }$, the quantum mechanics predicts that the measurement gives the
value $R^{\prime }$ of the quantity $\mathcal{R}$ with probability equal to $%
1$. It means that the $\delta $-like distribution is predicted, but not the
value $R^{\prime }$ of the measured quantity $\mathcal{R}$. To test the
prediction, we are to test whether the probability $w\left( R^{\prime
}\right) =1$ (but not whether the measured value of the quantity $\mathcal{R}
$ is equal to $R^{\prime }$). To test the distribution $w\left( R\right)
=\delta \left( R-R^{\prime }\right) $, we are to carry out a set of many $S$%
-measurements (i.e. $M$-measurement), but not a single measurement of the
quantity $\mathcal{R}$.

Influence of the $M$-measurement on the wave function $\psi $ of the
measured system (particle) is known as a reduction of the wave function. In
the case, when the unique value $R^{\prime }$ of the measured quantity $%
\mathcal{R}$ is obtained, the $M$-measurement is called a selective $M$%
-measurement (or $SM$-measurement). In general, we can obtain the unique
value $R^{\prime }$, only if the $M$-measurement is accompanied by a
selection. Only those elements of the statistical average particle $%
\left\langle \mathcal{S}\right\rangle $ (or the statistical ensemble $%
\mathcal{E}\left[ \mathcal{S}\right] $) are chosen, where the measured value
of the quantity $\mathcal{R}$ is equal to $R^{\prime }$. As a result the
chosen elements of $\left\langle \mathcal{S}\right\rangle $ form a new
statistical ensemble (or a new statistical average particle), whose state is
described by another wave function $\psi ^{\prime }$. Transformation $\psi
\rightarrow \psi ^{\prime }$ is a reduction of the wave function. This
reduction is conditioned by the selection of the statistical ensemble
elements. There is no mysticism in such a reduction, because its origin is
quite clear. But if we consider the selective quantum measurement as a
single measurement, where a selection is impossible, the reduction of the
wave function looks as a mystic procedure which accompanies the quantum
measurement. Thus, the statement that wave function describes the
statistical average particle $\left\langle \mathcal{S}\right\rangle $, (but
not the individual particle $\mathcal{S}$) is a crucial statement in the
explanation of the wave function reduction.

If in the $M$-measurement of the quantity $\mathcal{R}$ the measured value
is fixed by the measuring device, but the selection is not produced, the
reduction of the wave function is also takes place, but in this case the
reduction has another character. In this case the elements of the
statistical ensemble $\mathcal{E}\left[ \mathcal{S}\right] $, where the
measured value of the quantity $\mathcal{R}$ is $R_{i}$, form a statistical
subensemble $\mathcal{E}_{i}\left[ \mathcal{S}\right] $. These subensembles
evolve with different Hamiltonians depending on the value $R_{i}$. As a
result the statistical ensemble $\mathcal{E}\left[ \mathcal{S}\right] $,
whose state is described by the wave function $\psi $ turns to the set of
statistical ensembles $\mathcal{E}_{i}\left[ \mathcal{S}\right] $, taken
with the statistical weight (probability) $w\left( R_{i}\right) $. In this
case the wave function reduction leads to a transformation of the pure state
(wave function) into a mixed state (the density matrix). If instead of the $%
M $-measurement we consider a single measurement, the wave function
reduction appears to be a mystic procedure, because the selection cannot be
carried out in a single measurement. Again the statement that wave function
describes the statistical average particle $\left\langle \mathcal{S}%
\right\rangle $, (but not the individual particle $\mathcal{S}$) is a
crucial statement in the explanation of the wave function reduction.

The described properties of quantum experiment are not new for
representatives of the statistical interpretation \cite{B76,B70,B98,FP00},
but they looks as paradoxes for representatives of the Copenhagen
interpretation.

\appendix
\renewcommand{\theequation}{\Alph{section}.\arabic{equation}} %
\renewcommand{\thesection}{Appendix \Alph{section}.}

\section{Transformation of the action for the statistical ensemble.}

Let us transform the action (\ref{a1.22}) to the description in terms of the
function $\psi $. Instead of the independent variable $t=x^{0}$ we introduce
the variable $\xi _{0}$, and rewrite the action (\ref{a1.22}) in the form 
\begin{equation}
\mathcal{A}_{\mathcal{E}\left[ \mathcal{S}_{\mathrm{cl}}\right] }\left[ x%
\right] =\int \left\{ \frac{m\dot{x}^{\alpha }\dot{x}^{\alpha }}{2\dot{x}^{0}%
}\right\} d^{4}\xi ,\qquad \dot{x}^{k}\equiv \frac{\partial x^{k}}{\partial
\xi _{0}}  \label{A.1}
\end{equation}
where $\xi =\left\{ \xi _{0},\mathbf{\xi }\right\} =\left\{ \xi _{k}\right\} 
$,\ \ $k=0,1,2,3$, $x=\left\{ x^{k}\left( \xi \right) \right\} $,\ \ $%
k=0,1,2,3$. Here the variable $x^{0}$ is fictitious. Here and in what
follows, a summation over repeated Greek indices is produced $(1-3)$.

Let us consider variables $\xi =\xi \left( x\right) $ in (\ref{A.1}) as
dependent variables and variables $x$ as independent variables. Let the
Jacobian 
\begin{equation}
J=\frac{\partial \left( \xi _{0},\xi _{1},\xi _{2},\xi _{3}\right) }{%
\partial \left( x^{0},x^{1},x^{2},x^{3}\right) }=\det \left| \left| \xi
_{i,k}\right| \right| ,\qquad \xi _{i,k}\equiv \partial _{k}\xi _{i},\qquad
i,k=0,1,2,3  \label{A.3}
\end{equation}
be considered to be a multilinear function of $\xi _{i,k}$. Then 
\begin{equation}
d^{4}\xi =Jd^{4}x,\qquad \dot{x}^{i}\equiv \frac{\partial x^{i}}{\partial
\xi _{0}}\equiv \frac{\partial \left( x^{i},\xi _{1},\xi _{2},\xi
_{3}\right) }{\partial \left( \xi _{0},\xi _{1},\xi _{2},\xi _{3}\right) }%
=J^{-1}\frac{\partial J}{\partial \xi _{0,i}},\qquad i=0,1,2,3  \label{A.4}
\end{equation}
After transformation to dependent variables $\xi $ the action (\ref{A.1})
takes the form 
\begin{equation}
\mathcal{A}_{\mathcal{E}\left[ \mathcal{S}_{\mathrm{cl}}\right] }\left[ \xi %
\right] =\int \frac{m}{2}\frac{\partial J}{\partial \xi _{0,\alpha }}\frac{%
\partial J}{\partial \xi _{0,\alpha }}\left( \frac{\partial J}{\partial \xi
_{0,0}}\right) ^{-1}d^{4}x\mathbf{,}  \label{A.5}
\end{equation}

We introduce new variables 
\begin{equation}
j^{k}=\frac{\partial J}{\partial \xi _{0,k}},\qquad k=0,1,2,3  \label{A.6}
\end{equation}
by means of Lagrange multipliers $p_{k}$%
\begin{equation}
\mathcal{A}_{\mathcal{E}\left[ \mathcal{S}_{\mathrm{cl}}\right] }\left[ \xi
,j,p\right] =\int \left\{ \frac{m}{2}\frac{\partial J}{\partial \xi
_{0,\alpha }}\frac{\partial J}{\partial \xi _{0,\alpha }}\left( \frac{%
\partial J}{\partial \xi _{0,0}}\right) ^{-1}+p_{k}\left( \frac{\partial J}{%
\partial \xi _{0,k}}-j^{k}\right) \right\} d^{4}x\mathbf{,}  \label{A.7}
\end{equation}
Here and in what follows, a summation over repeated Latin indices is
produced $(0-3)$.

Note that according to (\ref{A.4}), the relations (\ref{A.6}) can be written
in the form 
\begin{equation}
j^{k}=\left\{ \frac{\partial J}{\partial \xi _{0,0}},\frac{\partial J}{%
\partial \xi _{0,0}}\left( J^{-1}\frac{\partial J}{\partial \xi _{0,\alpha }}%
\right) \left( J^{-1}\frac{\partial J}{\partial \xi _{0,0}}\right)
^{-1}\right\} =\left\{ \rho ,\rho \frac{dx^{\alpha }}{dt}\right\} ,\qquad
\rho \equiv \frac{\partial J}{\partial \xi _{0,0}}  \label{A.7a}
\end{equation}
It is clear from (\ref{A.7a}) that $j^{k}$ is the 4-flux of particles, with $%
\rho $ being its density.

Variation of (\ref{A.7}) with respect to $\xi _{i}$ gives 
\begin{equation}
\frac{\delta \mathcal{A}_{\mathcal{E}\left[ \mathcal{S}_{\mathrm{cl}}\right]
}}{\delta \xi _{i}}=-\partial _{l}\left( p_{k}\frac{\partial ^{2}J}{\partial
\xi _{0,k}\partial \xi _{i,l}}\right) =-\frac{\partial ^{2}J}{\partial \xi
_{0,k}\partial \xi _{i,l}}\partial _{l}p_{k}=0,\qquad i=0,1,2,3  \label{A.8}
\end{equation}
Using identities 
\begin{equation}
\frac{\partial ^{2}J}{\partial \xi _{0,k}\partial \xi _{i,l}}\equiv
J^{-1}\left( \frac{\partial J}{\partial \xi _{0,k}}\frac{\partial J}{%
\partial \xi _{i,l}}-\frac{\partial J}{\partial \xi _{0,l}}\frac{\partial J}{%
\partial \xi _{i,k}}\right)  \label{A.9}
\end{equation}
\begin{equation}
\frac{\partial J}{\partial \xi _{i,l}}\xi _{k,l}\equiv J\delta
_{k}^{i},\qquad \partial _{l}\frac{\partial ^{2}J}{\partial \xi
_{0,k}\partial \xi _{i,l}}\equiv 0  \label{A.10}
\end{equation}
one can test by direct substitution that the general solution of linear
equations (\ref{A.8}) has the form 
\begin{equation}
p_{k}=\frac{b}{2}\left( \partial _{k}\varphi +g^{\alpha }\left( \mathbf{\xi }%
\right) \partial _{k}\xi _{\alpha }\right) ,\qquad k=0,1,2,3  \label{A.11}
\end{equation}
where $b\neq 0$ is a constant, $g^{\alpha }\left( \mathbf{\xi }\right)
,\;\;\alpha =1,2,3$ are arbitrary functions of $\mathbf{\xi =}\left\{ \xi
_{1},\xi _{2},\xi _{3}\right\} $, and $\varphi $ is the dynamic variable $%
\xi _{0}$, which ceases to be fictitious. Note that it is the same
conceptual integration (\ref{b1.7}) which was discussed in Introduction. Let
us substitute (\ref{A.11}) in (\ref{A.7}). The term of the form $\partial
_{k}\varphi \partial J/\partial \xi _{0,k}$ is reduced to Jacobian and does
not contribute to dynamic equation. The terms of the form $\xi _{\alpha
,k}\partial J/\partial \xi _{0,k}$ vanish due to identities (\ref{A.10}). We
obtain 
\begin{equation}
\mathcal{A}_{\mathcal{E}\left[ \mathcal{S}_{\mathrm{cl}}\right] }\left[
\varphi ,\mathbf{\xi },j\right] =\int \left\{ \frac{m}{2}\frac{j^{\alpha
}j^{\alpha }}{j^{0}}-j^{k}p_{k}\right\} d^{4}x\mathbf{,}  \label{A.12}
\end{equation}
where quantities $p_{k}$ are determined by the relations (\ref{A.11})

Variation of the action (\ref{A.12}) with respect to $j^{k}$ gives 
\begin{equation}
p_{0}=-\frac{m}{2}\frac{j^{\alpha }j^{\alpha }}{\rho ^{2}},\qquad p_{\beta
}=m\frac{j^{\beta }}{\rho },\qquad \beta =1,2,3  \label{A.17}
\end{equation}

Now we eliminate the variables $\mathbf{j}=\left\{ j^{1},j^{2},j^{3}\right\} 
$ from the action (\ref{A.12}), using relation (\ref{A.17}). We obtain 
\begin{equation}
\mathcal{A}_{\mathcal{E}\left[ \mathcal{S}_{\mathrm{cl}}\right] }\left[ \rho
,\varphi ,\mathbf{\xi }\right] =-\int \left\{ p_{0}-\frac{p_{\beta }p_{\beta
}}{2m}\right\} \rho d^{4}x\mathbf{,}  \label{A.18}
\end{equation}
where $p_{k}$ is determined by the relation (\ref{A.11}).

Now instead of dependent variables $\rho ,\varphi ,\mathbf{\xi }$ we
introduce the $n$-component complex function $\psi $, defining it by
relations (\ref{s1.1}) -- (\ref{s5.5})

It is easy to verify that 
\begin{equation}
\rho =\psi ^{\ast }\psi ,\qquad \rho p_{0}\left( \varphi ,\mathbf{\xi }%
\right) =-\frac{ib}{2}(\psi ^{\ast }\partial _{0}\psi -\partial _{0}\psi
^{\ast }\cdot \psi )  \label{s5.6}
\end{equation}
\begin{equation}
\rho p_{\alpha }\left( \varphi ,\mathbf{\xi }\right) =-\frac{ib}{2}(\psi
^{\ast }\partial _{\alpha }\psi -\partial _{\alpha }\psi ^{\ast }\cdot \psi
),\qquad \alpha =1,2,3,  \label{s5.7}
\end{equation}
The variational problem with the action (\ref{A.18}) appears to be
equivalent to the variational problem with the action functional 
\begin{equation}
\mathcal{A}_{\mathcal{E}\left[ \mathcal{S}_{\mathrm{cl}}\right] }[\psi ,\psi
^{\ast }]=\int \left\{ \frac{ib}{2}(\psi ^{\ast }\partial _{0}\psi -\partial
_{0}\psi ^{\ast }\cdot \psi )-\frac{b^{2}}{8m}(\psi ^{\ast }\mathbf{\nabla }%
\psi -\mathbf{\nabla }\psi ^{\ast }\cdot \psi )^{2}\right\} \mathrm{d}^{4}x.
\label{s5.8}
\end{equation}
For the two-component function $\psi $ ($n=2$) the following identity takes
place 
\begin{equation}
(\mathbf{\nabla }\rho )^{2}-(\psi ^{\ast }\mathbf{\nabla }\psi -\mathbf{%
\nabla }\psi ^{\ast }\cdot \psi )^{2}\equiv 4\rho \mathbf{\nabla }\psi
^{\ast }\mathbf{\nabla }\psi -\rho ^{2}\sum\limits_{\alpha =1}^{\alpha
=3}\left( \mathbf{\nabla }s_{\alpha }\right) ^{2},  \label{s5.30}
\end{equation}
\begin{equation}
\rho \equiv \psi ^{\ast }\psi ,\qquad s\equiv \frac{\psi ^{\ast }\mathbf{%
\sigma }\psi }{\rho },\qquad \mathbf{\sigma }=\{\sigma _{\alpha }\},\qquad
\alpha =1,2,3,  \label{s5.31}
\end{equation}
where $\sigma _{\alpha }$ are the Pauli matrices. In virtue of the identity (%
\ref{s5.30}) the action (\ref{s5.8}) reduces to the form 
\begin{equation}
\mathcal{A}_{\mathcal{E}\left[ \mathcal{S}_{\mathrm{cl}}\right] }[\psi ,\psi
^{\ast }]=\int \left\{ \frac{ib}{2}\left( \psi ^{\ast }\partial _{0}\psi
-\partial _{0}\psi ^{\ast }\cdot \psi \right) -\frac{b^{2}}{2m}\mathbf{%
\nabla }\psi ^{\ast }\mathbf{\nabla }\psi +\frac{b^{2}}{8m}%
\sum\limits_{\alpha =1}^{\alpha =3}(\mathbf{\nabla }s_{\alpha })^{2}\rho +%
\frac{b^{2}}{8\rho m}(\mathbf{\nabla }\rho )^{2}\right\} \mathrm{d}^{4}x,
\label{s5.32}
\end{equation}
where $\mathbf{s}$ and $\rho $ are defined by the relations (\ref{s5.31}).

In the case of irrotational flow, when the two-component function $\psi $
has linear dependent components, for instance $\psi =\left\{ \psi
_{1},0\right\} $ the 3-vector $\mathbf{s}=$const, and the term containing
3-vector $\mathbf{s}$ vanishes. Then the action (\ref{s5.32}) for $\mathcal{E%
}\left[ \mathcal{S}_{\mathrm{cl}}\right] $ coincides with the action (\ref
{a1.16}) for $\mathcal{S}_{\mathrm{q}}$ with $\hbar =0$.

\end{document}